\documentclass[aps,superscriptaddress,onecolumn,preprintnumbers,showpacs,nofootinbib,preprint]{revtex4}
\usepackage{amssymb,amsmath,graphicx,dcolumn,bm,hyperref}

\begin{document}
\title{Isospin Symmetry Breaking in the Chiral Quark Model}

\newcommand*{\PKU}{School of Physics and State Key Laboratory of Nuclear Physics and
Technology, Peking University, Beijing 100871,
China}\affiliation{\PKU}
\newcommand*{\CHEP}{Center for High Energy
Physics, Peking University, Beijing 100871,
China}\affiliation{\CHEP}

\author{Huiying~Song}\affiliation{\PKU}
\author{Xinyu~Zhang}\affiliation{\PKU}
\author{Bo-Qiang~Ma\footnote{Corresponding author. Email address: \texttt{mabq@pku.edu.cn}}}\affiliation{\PKU}\affiliation{\CHEP}
\begin{abstract}
We discuss the isospin symmetry breaking (ISB) of the valence- and
sea-quark distributions between the proton and the neutron in the
framework of the chiral quark model. We assume that isospin symmetry
breaking is the result of mass differences between isospin
multiplets and then analyze the effects of isospin symmetry
breaking on the Gottfried sum rule and the NuTeV anomaly. We show
that, although both flavor asymmetry in the nucleon sea and the ISB
between the proton and the neutron can lead to the violation of the
Gottfried sum rule, the main contribution is from the flavor
asymmetry in the framework of the chiral quark model. We also find
that the correction to the NuTeV anomaly is in an opposite
direction, so the NuTeV anomaly cannot be removed by isospin
symmetry breaking in the chiral quark model. It is remarkable that
our results of ISB for both valence- and sea-quark distributions are
consistent with the Martin-Roberts-Stirling-Thorne
parametrization of quark distributions.
\end{abstract}

\maketitle

\section{Introduction}
Isospin symmetry was originally introduced to describe almost
identical properties of strong interaction of the proton and the
neutron by turning off their electromagnetic interaction, i.e.,
their charge information. This symmetry is commonly expected to be a
precise symmetry~\cite{Henley:1979ig,Miller:1990iz}, and its
breaking is assumed to be negligible in the phenomenological or
experimental analysis. This is, in general, true, since
electromagnetic interactions are weak compared with strong
interactions. However, it is possible for isospin symmetry breaking (ISB) to have
important influence on some experiments, especially
its effects on the parton distributions. Therefore, it is necessary
to analyze it carefully.

The isospin symmetry between the proton and the neutron originates
from the SU(2) symmetry between $u$ and $d$ quarks, which are
isospin doublets with isospin $I= 1/2$ and isospin three-components
($I_3$) 1/2 and -1/2, respectively. The isospin symmetry at parton
level indicates that the $u~(d,~\bar{u},~\bar{d})$-quark
distribution in the proton is equal to the $d~(u,~\bar{d},~\bar{u})$-quark distribution in the neutron. Accordingly, the ISBs of both valance-quark and sea-quark distributions are defined, respectively, as
\begin{eqnarray}
\delta u_{\mathrm{V}}(x)&=&u_{\mathrm{V}}^{\mathrm{p}}(x)-d_{\mathrm{V}}^{\mathrm{n}}(x),\nonumber\\
\delta
d_{\mathrm{V}}(x)&=&d_{\mathrm{V}}^{\mathrm{p}}(x)-u_{\mathrm{V}}^{\mathrm{n}}(x),\nonumber\\
\delta \bar{u}(x)&=&\bar{u}^{\mathrm{p}}(x)-\bar{d}^{\mathrm{n}}(x),\nonumber\\
\delta \bar{d}(x)&=&\bar{d}^{\mathrm{p}}(x)-\bar{u}^{\mathrm{n}}(x),
\end{eqnarray}
where
$q_{\mathrm{V}}^{\mathrm{N}}(x)=q^{\mathrm{N}}(x)-\bar{q}^{\mathrm{N}}(x)~(q=u,~d,~\mathrm{N}=\mathrm{p},~\mathrm{n}).$

ISB at the parton level and its possible consequences for several
processes were first investigated by one of us~\cite{Ma:1991ac}. It
was pointed out that both flavor asymmetry in the nucleon sea and
isospin symmetry breaking between the proton and the neutron can
lead to the violation of the Gottfried sum rule reported by the New
Muon Collaboration~\cite{Amaudruz:1991at,Arneodo:1994sh}. The
possibility of distinguishing these two effects was also discussed
in detail~\cite{Ma:1992gp}.

In 2002, the NuTeV Collaboration~\cite{Zeller:2001hh} extracted
$\sin^{2}\theta_{\mathrm{W}}$ by measuring the ratios of neutral
current to charged current $\nu$ and $\bar{\nu}$ cross sections on
iron targets. The reported
$\sin^{2}\theta_{\mathrm{W}}=0.2277\pm0.0013\left(\mathrm{stat}\right)\pm0.0009\left(\mathrm{syst}\right)$
has approximately 3 standard deviations above the world average
value $\sin^{2}\theta_{\mathrm{W}}=0.2227\pm0.0004$ measured in
other electroweak processes. This remarkable deviation is called the
NuTeV anomaly and was discussed in a number of papers from various
aspects, including new physics beyond the standard
model~\cite{Davidson:2001ji}, the nuclear
effect~\cite{Kovalenko:2002xe}, nonisoscalar
targets~\cite{Kumano:2002ra}, and strange-antistrange
asymmetry~\cite{Cao:2003ny,Ding:2004ht,Ding:2004dv}. Moreover, the
possible influence of ISB on this measurement was also studied in a
series of
papers~\cite{Sather:1991je,Rodionov:1994cg,Davidson:1997mb,Londergan:2003ij,Cao:2000dj,Gluck:2005xh,Ding:2006ud}.
However, the correction from ISB to the NuTeV anomaly is still not
conclusive.

The Martin-Roberts-Stirling-Thorne (MRST) group~\cite{Martin:2003sk}
provided some evidence to support the ISB effects on parton
distributions of both valance and sea quarks and included ISB in
the parametrization based on experimental data. They obtained the ISB of
valance quarks as
\begin{eqnarray}
\delta u_{\mathrm{V}}=-\delta
d_{\mathrm{V}}=\kappa(1-x)^{4}x^{-0.5}(x-0.0909),
\end{eqnarray}
where $-0.8\leq\kappa\leq+0.65$ with a $90\%$ confidence level, and
the best fit value is $\kappa=-0.2$. They also obtained the ISB of sea
quarks, as can be deduced from Eqs.~(28) and (29) in
Ref.~\cite{Martin:2003sk},
\begin{eqnarray}
\delta\bar{u}(x)=k\bar{u}^{\mathrm{p}}(x),~~~~\delta\bar{d}(x)=k\bar{d}^{\mathrm{p}}(x),
\end{eqnarray}
with the best fit value $k=0.08$.

In this paper, we calculate the ISB of the valance- and sea-quark
distributions between the proton and the neutron in the chiral quark
model and discuss some possible effects of ISB.  We assume that the
ISB between the proton and the neutron is entirely from the mass
difference between isospin multiplets at both hadron and parton
levels.\footnote{As mass difference between isospin multiplets,
especially that between $u$ and $d$ quarks, is not entirely due to
charge difference, we refer such effect as Isospin Symmetry Breaking
(ISB) instead of Charge Symmetry Breaking (CSB) as called in some
papers.} In Sec.~\ref{section2}, we compute ISB in the chiral quark
model, with the constituent-quark-model results as the bare
constituent-quark-distribution inputs. Then, we calculate the ISB
effect on the violation of the Gottfried sum rule. In
Sec.~\ref{section3}, we discuss the ISB correction to the measurement of the
weak angle and point out the significant influence on the NuTeV
anomaly. In Sec.~\ref{section4}, we provide summaries of the paper.

\section{isospin symmetry breaking in the chiral quark model}\label{section2}
The chiral quark model, established by
Weinberg~\cite{Weinberg:1978kz} and developed by Manohar and
Georgi~\cite{Manohar:1983md}, has an apt description of its
important degrees of freedom in terms of quarks, gluons, and
Goldstone~(GS) bosons at momentum scales relating to hadron
structure. This model is successful in explaining numerous problems,
including the violation of the Gottfried sum rule from the aspect
of flavor asymmetry in the nucleon
sea~\cite{Eichten:1991mt,Wakamatsu:1991tj}, the proton spin
crisis~\cite{Ashman:1987hv,Cheng:1994zn,Song:1997bp}, and the NuTeV
anomaly resulting from the strange-antistrange
asymmetry~\cite{Ding:2004dv}, and has been widely recognized as an
effective theory of QCD at the low-energy scale.

In the chiral quark model, the minor effects of the internal gluons are
negligible. The valence quarks contained in the nucleon fluctuate
into quarks plus GS bosons, which spontaneously break chiral
symmetry. Then, the effective interaction Lagrangian is
\begin{equation}
L=\bar{\psi}\left(iD_{\mu}+V_{\mu}\right)\gamma^{\mu}\psi+ig_{\mathrm{A}}\bar{\psi}A_{\mu}\gamma^{\mu}\gamma_{5}\psi+\cdots,
\end{equation}
where
\begin{equation}
\psi=\left(%
\begin{array}{c}
  u \\
  d \\
  s \\
\end{array}%
\right)
\end{equation}
is the quark field and $D_{\mu}=\partial_{\mu}+igG_{\mu}$ is the
gauge-covariant derivative of QCD. $G_{\mu}$ stands for the gluon
field, $g$ stands for the strong coupling constant, and $g_{\mathrm{A}}$
stands for the axial-vector coupling constant. $V_{\mu}$ and
$A_{\mu}$ are the vector and the axial-vector currents, which are
defined as
\begin{equation}
\left(%
\begin{array}{c}
  V_{\mu} \\
  A_{\mu} \\
\end{array}%
\right)=\frac{1}{2}\left(\xi^{+}\partial_{\mu}\xi\pm\xi\partial_{\mu}\xi^{+}\right),
\end{equation}
where $\xi=\mathrm{exp}(i\Pi/f)$, and $\Pi$  has the form:
\begin{equation}
\Pi\equiv\frac{1}{\sqrt{2}}\left(
\begin{array}{ccc}
  \frac{\pi^{0}}{\sqrt{2}}+\frac{\eta}{\sqrt{6}} & \pi^{+} & K^{+} \\
  \pi^{-} & -\frac{\pi^{0}}{\sqrt{2}}+\frac{\eta}{\sqrt{6}} & K^{0} \\
  K^{-} & \overline{K^{0}} & \frac{-2\eta}{\sqrt{6}} \\
\end{array}
\right).
\end{equation}
Expanding $V_{\mu}$ and $A_{\mu}$ in powers of $\Pi/f$, one gets
$V_{\mu}=0+O(\Pi/f)^{2}$ and
$A_{\mu}=i\partial_{\mu}\Pi/f+O(\Pi/f)^{2}$. The pseudoscalar decay
constant is $f\simeq93$~MeV. Thus, the effective interaction
Lagrangian between GS bosons and quarks in the leading order
becomes~\cite{Eichten:1991mt}
\begin{equation}
L_{\Pi
q}=-\frac{g_{\mathrm{A}}}{f}\bar{\psi}\partial_{\mu}\Pi\gamma^{\mu}\gamma_{5}\psi.
\end{equation}
Based on the time-ordered perturbative theory in the infinite
momentum frame, all particles are on-mass-shell, and the
factorization of the subprocess is automatic, so we can express the
quark distributions inside a nucleon as a convolution of a
constituent-quark distribution in a nucleon and the structure
functions of a constituent quark. Since the $\eta$ is relatively
heavy, we neglect the minor contribution from its suppressed
fluctuation in this paper. Then, the light-front Fock decompositions
of constituent-quark wave functions are
\begin{equation}
|U\rangle=\sqrt{Z_{u}}|u_{0}\rangle+a_{\pi^{+}}|d\pi^{+}\rangle+\frac{a_{u\pi^{0}}}{%
\sqrt{2}}|u\pi^{0}\rangle+a_{K^{+}}|sK^{+}\rangle,\label{u}
\end{equation}
\begin{equation}
|D\rangle=\sqrt{Z_{d}}|d_{0}\rangle+a_{\pi^{-}}|u\pi^{-}\rangle+\frac{a_{d\pi^{0}}}{%
\sqrt{2}}|d\pi^{0}\rangle+a_{K^{0}}|sK^{0}\rangle,\label{d}
\end{equation}
where $Z_{u}$ and $Z_{d}$ are the renormalization constants for the
bare constituent $u$ quark $|u_{0}\rangle$ and $d$ quark
$|d_{0}\rangle$, respectively, and $|a_{\alpha}|^{2}$ ($\alpha=\pi,
K$) are the probabilities to find GS bosons in the dressed
constituent-quark states $|U\rangle$ and $|D\rangle$.  In the chiral
quark model, the fluctuation of a bare constituent quark into a GS
boson and a recoil bare constituent quark is given
as~\cite{Suzuki:1997wv}
\begin{equation}
q_{j}(x)=\int^{1}_{x}\frac{\textmd{d}y}{y}P_{j\alpha/i}(y)q_{i}\left(\frac{x}{y}\right),\label{q}
\end{equation}
where $P_{j\alpha/i}(y)$ is the splitting function, which gives the
probability of finding a constituent quark $j$ carrying the
light-cone momentum fraction $y$ together with a spectator GS
boson~$\alpha$,
\begin{eqnarray}
P_{j\alpha/i}(y)=\frac{1}{8\pi^{2}}\left(\frac{g_{\mathrm{A}}\overline{m}}{f}\right)^{2}\int
\textmd{d}k^{2}_{T}\frac{(m_{j}-m_{i}y)^{2}+k^{2}_{T}}{y^{2}(1-y)[m_{i}^{2}-M^{2}_{j\alpha}]^{2}}.
\label{splitting}
\end{eqnarray}
$m_{i}$, $m_{j}$, and $m_{\alpha}$ are the masses of the $i$- and $
j$-constituent quarks and the pseudoscalar meson $\alpha$,
respectively, and $\overline{m}=(m_{i}+m_{j})/2$ is the average mass
of the constituent quarks.
$M^{2}_{j\alpha}=\left(m^{2}_{j}+k^{2}_{T}\right)/y+\left(m^{2}_{\alpha}+k^{2}_{T}\right)/\left(1-y\right)$
is the square of the invariant mass of the final state. We can also
write the internal structure of GS bosons in the following form
\begin{equation}
q_{k}(x)=\int\frac{\textmd{d}y_{1}}{y_{1}}\frac{\textmd{d}
y_{2}}{y_{2}}V_{k/\alpha}\left(\frac{x}{y_{1}}\right)P_{\alpha
j/i}\left(\frac{y_{1}}{y_{2}}\right)q_{i}\left(y_{2}\right),
\end{equation}
where $V_{k/\alpha}(x)$ is the quark $k$ distribution function in
$\alpha$ and satisfies the normalization
$\int_{0}^{1}V_{k/\alpha}(x)dx=1$.

When we take ISB into consideration, the renormalization constant
$Z$ should take the form
\begin{eqnarray}
Z_{u}=1-\langle P_{\pi^{+}}\rangle-\frac{1}{2}\langle
P_{u\pi^{0}}\rangle-\langle P_{K^{+}}\rangle,\nonumber\\
Z_{d}=1-\langle P_{\pi^{-}}\rangle-\frac{1}{2}\langle
P_{d\pi^{0}}\rangle-\langle P_{K^{0}}\rangle,
\end{eqnarray}
where $\langle P_{\alpha}\rangle\equiv \langle
P_{j\alpha/i}\rangle=\langle P_{\alpha
j/i}\rangle=\int^{1}_{0}x^{n-1}P_{j\alpha/i}(x)\mathrm{d}x$~\cite{Suzuki:1997wv}.
It is conventional to specify the momentum cutoff function at the
quark-GS-boson vertex as
\begin{equation}
g_{\mathrm{A}}\rightarrow
g_{\mathrm{A}}^{\prime}\textmd{exp}\bigg{[}\frac{m^{2}_{i}-M^{2}_{j\alpha}}{4\Lambda^{2}}\bigg{]},
\end{equation}
where $g_{\mathrm{A}}^{\prime}=1$, following the large $N_{c}$
argument~\cite{Weinberg:1990xm}, and $\Lambda$ is the cutoff
parameter, which is determined by the experimental data of the
Gottfried sum and the constituent-quark-mass inputs for the pion.
Such a form factor has the correct $t$- and $u$-channel symmetry,
$P_{j \alpha /i} (y) = P_{\alpha j/i} (1-y)$. Then, one can obtain
the quark-distribution functions in the proton~\cite{Suzuki:1997wv},
\begin{eqnarray}
u(x)&=&Z_{u}u_{0}(x)+P_{u\pi^{-}/d}\otimes
d_{0}(x)+V_{u/\pi^{+}}\otimes P_{\pi^{+}d/u}\otimes
u_{0}(x)+\frac{1}{2}P_{u\pi^{0}/u}\otimes
u_{0}(x)\nonumber\\&+&V_{u/K^{+}}\otimes P_{K^{+}s/u}\otimes
u_{0}(x)+
\frac{1}{2}V_{u/\pi^{0}}\otimes \left[P_{\pi^{0}u/u}\otimes u_{0}(x)+P_{\pi^{0}d/d}\otimes d_{0}(x)\right],\nonumber\\
d(x)&=&Z_{d}d_{0}(x)+P_{d\pi^{+}/u}\otimes
u_{0}(x)+V_{d/\pi^{-}}\otimes P_{\pi^{-}u/d }\otimes d_{0}(x)+
\frac{1}{2}P_{d\pi^{0}/d}\otimes
d_{0}(x)\nonumber\\&+&V_{d/K^{0}}\otimes P_{K^{0}s/d}\otimes
d_{0}(x) +\frac{1}{2}V_{d/\pi^{0}}\otimes \left[P_{\pi^{0}u/u }\otimes
u_{0}(x)+P_{\pi^{0}d/d}\otimes d_{0}(x)\right],\nonumber\\
\bar{u}(x)&=&V_{\bar{u}/\pi^{-}}\otimes P_{\pi^{-}u/d}\otimes
d_{0}(x)+\frac{1}{2}V_{\bar{u}/\pi^{0}}\otimes
\left[P_{\pi^{0}u/u}\otimes
u_{0}(x)+P_{\pi^{0}d/d}\otimes d_{0}(x)\right],\nonumber\\
\bar{d}(x)&=&V_{\bar{d}/\pi^{+}}\otimes P_{\pi^{+}d/u}\otimes
u_{0}(x)+\frac{1}{2}V_{\bar{d}/\pi^{0}}\otimes
\left[P_{\pi^{0}u/u}\otimes u_{0}(x)+P_{\pi^{0}d/d}\otimes d_{0}(x)\right],
\end{eqnarray}
where the constituent quark-distributions $u_{0}$ and $d_{0}$ are
normalized to two and one, respectively. Convolution integrals are
defined as
\begin{eqnarray}
P_{j \alpha / i}\otimes q_i &=&\int_{x}^{1}\frac{\textmd{d}y}{y}P_{j
\alpha / i}\left(y\right)q_i\left(\frac{x}{y}\right),\nonumber\\
V_{k/ \alpha}\otimes P_{\alpha j/i}\otimes
q_i&=&\int_{x}^{1}\frac{\textmd{d}y_{1}}{y_{1}}\int_{y_{1}}^{1}\frac{\textmd{d}y_{2}}{y_{2}}V_{k/
\alpha}\left( \frac{x}{y_{1}}\right)P_{\alpha
j/i}\left(\frac{y_{1}}{y_{2}}\right)q_{i}\left(y_{2}\right).
\end{eqnarray}
In addition, $V_{k/\alpha}(x)$ follows the relationship
\begin{eqnarray}
&&V_{u/\pi^{+}}=V_{\bar{d}/\pi^{+}}=V_{d/\pi^{-}}=V_{\bar{u}/\pi^{-}}=2V_{u/\pi^{0}}
=2V_{\bar{u}/\pi^{0}}=2V_{d/\pi^{0}}=2V_{\bar{d}/\pi^{0}}
=\frac{1}{2}V_{\pi},\nonumber\\
&&V_{u/K^{+}}=V_{d/K^{0}}.
\end{eqnarray}

We postulate that the bare-quark distributions are isospin-symmetric
between the proton and the neutron, so we can obtain the quark
distributions of the neutron by interchanging $u_{0}$ and $d_{0}$.
Employing the quark distributions of the chiral quark model, we get
the Gottfried sum determined by the difference between the proton
and the neutron structure functions,
\begin{eqnarray}
S_{\mathrm{G}}&=&\int^{1}_{0}\frac{\mathrm{d}x}{x}\left[F^{\mathrm{p}}_{2}(x)-F^{\mathrm{n}}_{2}(x)\right]\nonumber\\
&=&\frac{1}{9}\int_{0}^{1}\mathrm{d}x\left[4u^{\mathrm{p}}(x)+4\bar{u}^{\mathrm{p}}(x)-4u^{\mathrm{n}}(x)-4\bar{u}^{\mathrm{n}}(x)+d^{\mathrm{p}}(x)+\bar{d}^{\mathrm{p}}(x)-d^{\mathrm{n}}(x)-\bar{d}^{\mathrm{n}}(x)\right]\nonumber\\&
=&\frac{1}{3}+\int^{1}_{0}\mathrm{d}x\left\{\frac{8}{9}\left[\bar{u}^{\mathrm{p}}(x)-\bar{u}^{\mathrm{n}}(x)\right]+\frac{2}{9}\left[\bar{d}^{\mathrm{p}}(x)-\bar{d}^{\mathrm{n}}(x)\right]\right\}\nonumber\\
&=&\frac{1}{3}-\frac{8}{9}\left<P_{\pi^{-}}\right>+\frac{2}{9}\left<P_{\pi^{+}}\right>+\frac{5}{18}
\left(\left<P_{u\pi^{0}}\right>-\left<P_{d\pi^{0}}\right>\right).\label{gottfried}
\end{eqnarray}

We assume that the ISB is entirely from the mass difference between
isospin multiplets. In this paper, we adopt
$(m_{u}+m_{d})/2=330$~MeV, $m_{\pi^{\pm}}=139.6$~MeV,
$m_{\pi^{0}}=135$~MeV, $m_{K^{\pm}}=493.7$~MeV, and
$m_{K^{0}}=497.6$~MeV. We choose two sets of the mass difference
between $u$ and $d$ quarks, namely $\delta m=4$~MeV and $\delta
m=8$~MeV, respectively, in order to show the dependence on this
important parameter. Based on Eq.~(\ref{gottfried}) and the
experimental data of the Gottfried sum~\cite{Arneodo:1994sh}, one
can find that the appropriate value for $\Lambda_{\pi}$ is
$1500$~MeV. However, one cannot determine $\Lambda_{K}$ in the same
method, because $\langle P_{K}\rangle$ in the Gottfried sum is
canceled out. Usually, it is assumed that $\Lambda_{K}=\Lambda_{\pi}=1500~\mathrm{MeV}$\cite{Suzuki:1997wv,Szczurek:1996tp}.
However, it is implied by the SU$\left(3\right)_{f}$ symmetry
breaking that $\left<P_{K}\right>$ should be smaller, and,
accordingly, one should adopt a smaller $\Lambda_{K}$. In this paper,
we adopt a wide range of $\Lambda_{K}$ from $900$ to $1500$~MeV. In
addition, the parton distributions of mesons are the parametrization GRS98 given by Gluck-Reya-Stratmann~\cite{Gluck:1997ww}, since the parametrization is
more approximate to the actual value,
\begin{eqnarray}
V_{\pi}(x)=0.942x^{-0.501}(1+0.632\sqrt{x})(1-x)^{0.367},\nonumber\\
V_{u/K^{+}}(x)=V_{d/K^{0}}(x)=0.541(1-x)^{0.17}V_{\pi}(x).
\end{eqnarray}
We should point out that, in principle, it is possible that the parton
distributions of different mesons in the same multiplet are
different, and this can contribute to ISB simultaneously. However,
in this paper, we simply neglect this possibility, and calculations
in future can be improved if we have a better understanding of the
quark structure of mesons. Moreover, we have to specify constituent-quark distributions $u_{0}$ and $d_{0}$, but there is no proper
parametrization of them because they are not directly related to
observable quantities in experiments. In this paper, we adopt the
constituent-quark-model distributions~\cite{Hwa:1980mv} as inputs
for constituent-quark distributions. For the proton, we have
\begin{eqnarray}
u_{0}(x)&=&\frac{2x^{c_{1}}(1-x)^{c_{1}+c_{2}+1}}{\textmd{B}[c_{1}+1,c_{1}+c_{2}+2]},\nonumber\\
d_{0}(x)&=&\frac{x^{c_{2}}(1-x)^{2c_{1}+1}}{\textmd{B}[c_{2}+1,2c_{1}+2]},
\end{eqnarray}
where $\textmd{B}[i,j]$ is the Euler beta function. Such distributions
satisfy the number and the momentum sum rules
\begin{eqnarray}
\int^{1}_{0}u_{0}(x)\textmd{d}x=2, ~~~~
\int^{1}_{0}d_{0}(x)\textmd{d}x=1,\nonumber\\
\int_{0}^{1}xu_{0}(x)\textmd{d}x+\int_{0}^{1}xd_{0}(x)\textmd{d}x=1.
\end{eqnarray}
$c_{1}=0.65$ and $c_{2}=0.35$ are adopted in the calculation, following the original choice~\cite{Hwa:1980mv,Kua:1999yf}.

\begin{figure}
\begin{center}
\includegraphics[width=0.95\textwidth]{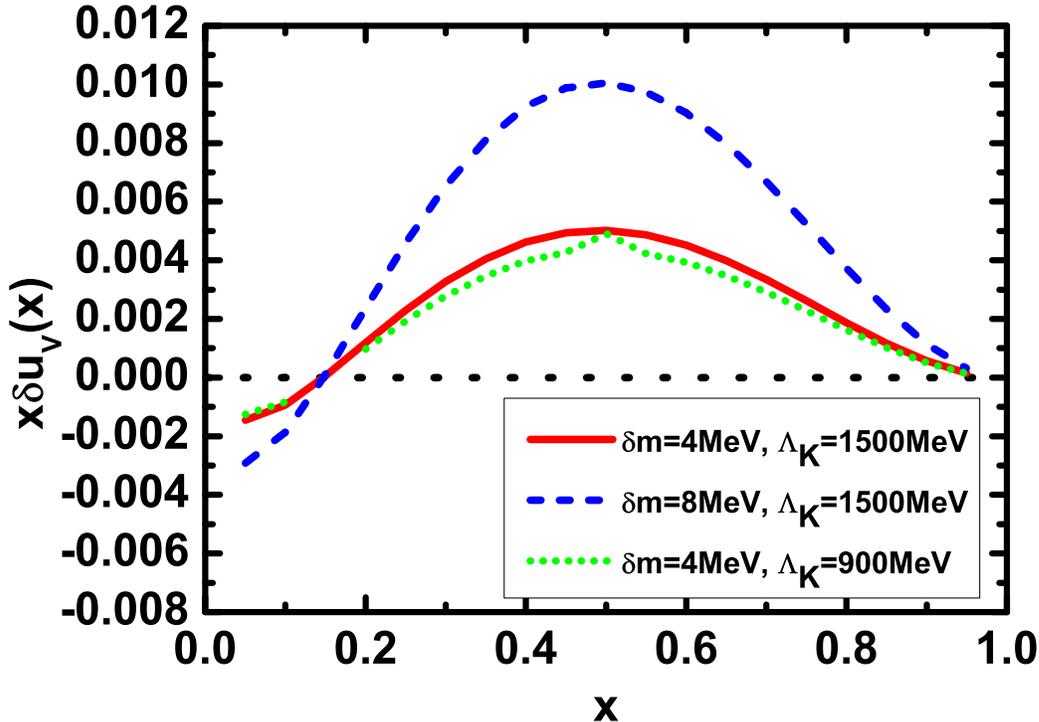}
\caption{\small The ISB of the $u_{\mathrm{V}}$-quark distribution
$x\delta u_{\mathrm{V}}(x)$ versus $x$ in the chiral quark model
with different inputs. The red solid line is the result with $\delta
m=4$~MeV and $\Lambda_K=1500$~MeV as inputs. The blue dashed line is
the result with $\delta m=8$~MeV and $\Lambda_K=1500$~MeV as inputs.
The green dotted line is the result with $\delta m=4$~MeV and $\Lambda_K=900$~MeV as inputs.}\label{uvalance}
\end{center}
\end{figure}

\begin{figure}
\begin{center}
\includegraphics[width=0.95\textwidth]{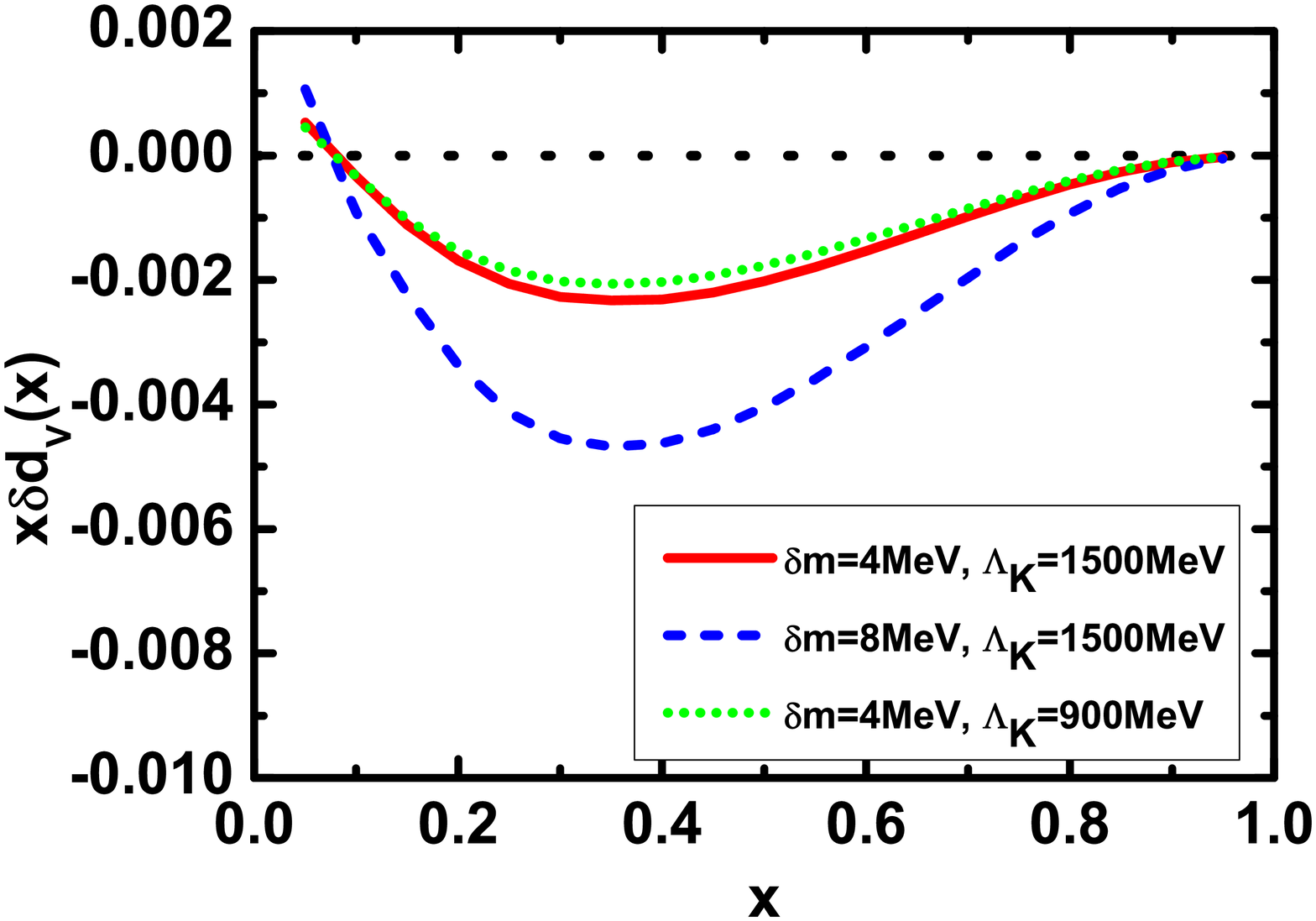}
\caption{\small The ISB of the $d_{\mathrm{V}}$-quark distribution
$x\delta d_{\mathrm{V}}(x)$ versus $x$ in the chiral quark model
with different inputs. The red solid line is the result with $\delta
m=4$~MeV and $\Lambda_K=1500$~MeV as inputs. The blue dashed line is
the result with $\delta m=8$~MeV and $\Lambda_K=1500$~MeV as inputs.
The green dotted line is the result with $\delta m=4$~MeV and
$\Lambda_K=900$~MeV as inputs.}\label{dvalance}
\end{center}
\end{figure}

\begin{figure}
\begin{center}
\includegraphics[width=0.95\textwidth]{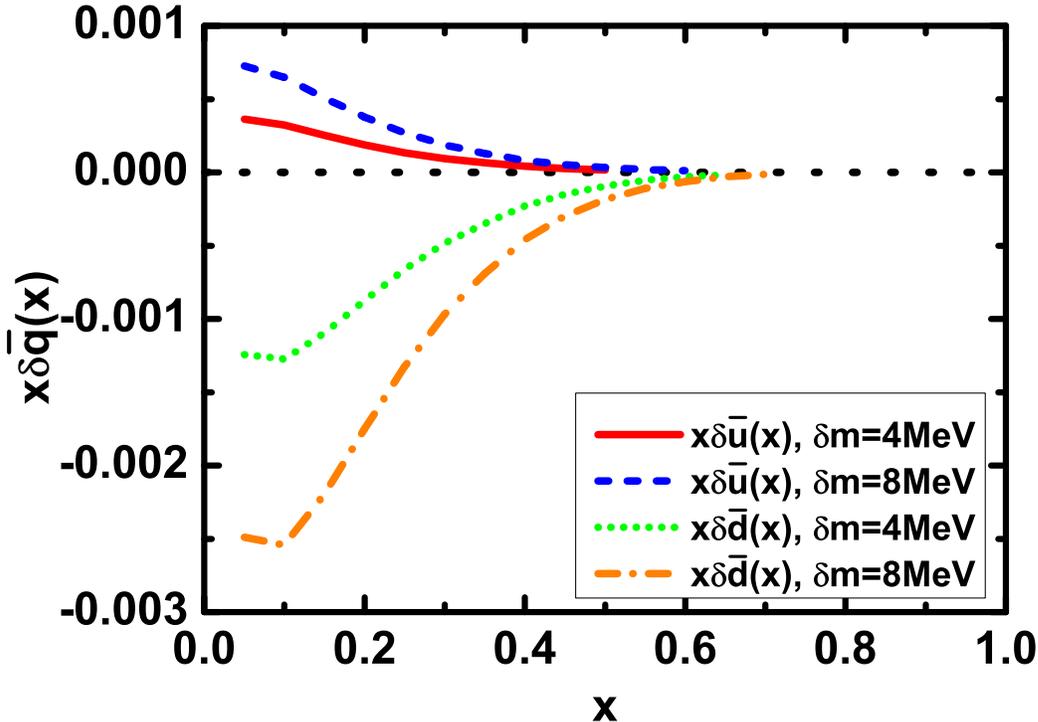}
\caption{\small The ISB of the sea-quark distributions $x\delta
\bar{q}(x)$ versus $x$ in the chiral quark model. The red solid line
and the blue dashed line are the behaviors of $x\delta \bar{u}(x)$,
with $\delta m=4$~MeV and $\delta m=8$~MeV, respectively. The green
dotted line and the orange dash-dotted line are the behaviors of
$x\delta \bar{d}(x)$, with $\delta m=4$~MeV and $\delta m=8$~MeV, respectively.}\label{sea}
\end{center}
\end{figure}

We display the ISB of the valance- and sea-quark distributions in
Figs.~\ref{uvalance}, \ref{dvalance}, and \ref{sea}, respectively. It
is shown that in most regions, $x\delta u_\mathrm{V}(x)>0$ and
$x\delta \bar{u}(x)>0$, and on the contrary that $x\delta
d_\mathrm{V}(x)<0$ and $x\delta \bar{d}(x)<0$. Our predictions that
$x\delta \bar{u}(x)>0$ and $x\delta \bar{d}(x)<0$ are consistent
with the MRST parametrization~\cite{Martin:2003sk}, and, moreover,
the shapes of $x\delta \bar{u}(x)$ and $x\delta \bar{d}(x)$ are
similar to the best phenomenological fitting results given by the
MRST group. We should point out that our results are analogous to
the results calculated in the framework of the meson cloudy model by Cao
and Signal~\cite{Cao:2000dj}, and the shapes and magnitudes of
$x\delta \bar{u}(x)$ and $x\delta \bar{d}(x)$ are similar to the
results given in the framework of the radiatively generated
ISB~\cite{Gluck:2005xh}, but with different signs. It can also be
found that the difference between various choices of $\Lambda_K$ is
minor, but the different choices of $\delta m$ can have remarkable
influence on the distributions. Especially, larger $\delta m$ can
lead to larger ISB, and this is concordant with our principle that
ISB results from the mass difference between isospin multiplets at
both hadron and parton levels. From the figures, we can see that
$\delta u_\mathrm{V}(x)$ reaches a maximum value at $x\approx 0.5$,
and $\delta d_\mathrm{V}(x)$ has a minimum value at $x\approx 0.4$.
It should also be noted that $\delta q_\mathrm{V}(x)$ $(q=u,d)$ must
have at least one zero point due to the valance-quark-normalization
conditions. We should also point out that at large $x$, $\delta
u_\mathrm{V}/u_\mathrm{V} \approx -\delta
d_\mathrm{V}/d_\mathrm{V}$, and this implies that the magnitudes of
the ISB for $u_{\mathrm{V}}$ and $d_{\mathrm{V}}$ are almost the
same, but with opposite signs. Moreover, although both flavor
asymmetry in the nucleon sea and the ISB between the proton and the
neutron can lead to the violation of the Gottfried sum rule, the
main contribution is from the flavor asymmetry in the framework of
the chiral quark model.

\section{The contribution from isospin symmetry breaking to the NuTeV anomaly}\label{section3}
The measured $\sin^2\theta_\mathrm{W}$ by the NuTeV Collaboration is closely
related to the Paschos-Wolfenstein (PW) ratio~\cite{Paschos:1972kj}
\begin{eqnarray}
R^{-}=\frac{\left<\sigma_{\mathrm{NC}}^{\nu
\mathrm{N}}\right>-\left<\sigma_{\mathrm{NC}}^{\overline{\nu}\mathrm{N}}\right>}{\left<\sigma_{\mathrm{CC}}^{\nu
\mathrm{N}}\right>-\left<\sigma_{\mathrm{CC}}^{\overline{\nu}\mathrm{N}}\right>}=\frac{1}{2}-\sin^{2}\theta_{\mathrm{W}},\label{pw}
\end{eqnarray}
where $\left<\sigma_{\mathrm{NC}}^{\nu \mathrm{N}}\right>$ is the
neutral-current-inclusive cross section for a neutrino on an isoscalar
target. If we take the ISB between the proton and the neutron into
account, we obtain
\begin{eqnarray}
R^{-}_{\mathrm{N}}=\frac{\left<\sigma_{\mathrm{NC}}^{\nu
\mathrm{N}}\right>-\left<\sigma_{\mathrm{NC}}^{\overline{\nu}\mathrm{N}}\right>}{\left<\sigma_{\mathrm{CC}}^{\nu
\mathrm{N}}\right>-\left<\sigma_{\mathrm{CC}}^{\overline{\nu}\mathrm{N}}\right>}
=R^{-}+\delta R^{\mathrm{ISB}}_{\mathrm{PW}},\label{mpw}
\end{eqnarray}
where $\delta R^{\mathrm{ISB}}_{\mathrm{PW}}$ is the correction from the
ISB to the PW ratio and takes the form
\begin{eqnarray}
\delta
R^{\mathrm{ISB}}_{\mathrm{PW}}=\bigg{(}\frac{1}{2}-\frac{7}{6}\sin^{2}\theta_{\mathrm{W}}\bigg{)}\frac{\int^{1}_{0}x\bigg{[}\delta
u_{\mathrm{V}}(x)-\delta d_{\mathrm{V}}(x)\bigg{]}\mathrm{d}x}
{\int^{1}_{0}x\bigg{[}u_{\mathrm{V}}(x)+d_{\mathrm{V}}(x)\bigg{]}\mathrm{d}x},\label{rs}
\end{eqnarray}
with $u_{\mathrm{V}}(x)$ and $d_{\mathrm{V}}(x)$ standing for
valance-quark distributions of the proton. We show the
renormalization constant $Z$, the total momentum fraction of valance
quarks
$Q_{\mathrm{V}}=\int^{1}_{0}x\left[u_{\mathrm{V}}(x)+d_{\mathrm{V}}(x)\right]\mathrm{d}x$,
and the correction of the ISB to the NuTeV anomaly $\Delta
R^{\mathrm{ISB}}_{\mathrm{PW}}$, with different $\delta m$ and
$\Lambda_{K}$ as inputs in Table~\ref{ISB}. It can be found that the ISB
correction is of the order of magnitude of $10^{-3}$ and is more
significant with a larger $\delta m$ or $\Lambda_K$. Our result is
consistent with the range $-0.009\leq\Delta
R_{\mathrm{PW}}^{\mathrm{ISB}}\leq+0.007$, which is derived based on
the parametrization given by the MRST group~\cite{Martin:2003sk}. We
should stress that the correction is remarkable, since the NuTeV
anomaly can be totally removed if $\Delta R_{\mathrm{PW}}=-0.005$,
and, consequently, we should pay special attention to ISB in such
problem. It is also worthwhile to point out that the correction is
in an opposite direction to remove the NuTeV anomaly in the chiral
quark model. Such a conclusion is the same as that given in the
baryon-meson fluctuation model~\cite{Ding:2006ud}, but the value is
one or 2 orders of magnitude larger. Our result of the ISB
correction to the NuTeV anomaly differs from the results in
Refs.~\cite{Londergan:2003ij,Gluck:2005xh}.

\begin{table}[!htbp]
\caption{\small The renormalization constant, the total momentum
fraction of valance quarks, and the correction of the ISB to the NuTeV
anomaly in the chiral quark model.}\label{ISB}
\begin{center}
\begin{tabular}{cccccc}
  \hline
  \hline
  $\delta m$~(MeV)& $\Lambda_{K}$~(MeV) & $Z_{u}$ & $Z_{d}$ & $Q_{\mathrm{V}}$&$\Delta R^{\mathrm{ISB}}_{\mathrm{PW}}$\\
  $4$ & $900$  &  $0.7497$  & $0.7463$ & $0.8451$ & $0.0008$  \\
  $4$ & $1200$  &  $0.7220$  & $0.7185$ & $0.8222$ & $0.0008$  \\
  $4$ & $1500$  &  $0.6932$  & $0.6896$ & $0.7985$ & $0.0009$  \\
  $8$ & $900$  &  $0.7515$  & $0.7444$ & $0.8455$ & $0.0016$  \\
  $8$ & $1200$  &  $0.7239$  & $0.7165$ & $0.8227$ & $0.0017$  \\
  $8$ & $1500$  &  $0.6953$  & $0.6874$ & $0.7990$ & $0.0019$  \\
  \hline
  \hline
\end{tabular}
\end{center}
\end{table}

\section{summary}\label{section4}
In this paper, we discuss the ISB of the
valance-quark and the sea-quark distributions between the proton and
the neutron in the framework of the chiral quark model. We assume
that isospin symmetry breaking is the result of mass differences
between isospin multiplets. Then, we analyze the effects of isospin
symmetry breaking on the Gottfried sum rule and the NuTeV anomaly.
We show that, although both flavor asymmetry in the nucleon sea and
the ISB between the proton and the neutron can lead to the violation
of the Gottfried sum rule, the main contribution is from the flavor
asymmetry in the framework of the chiral quark model. It is
remarkable that our results of ISB for both the valence-quark and
sea-quark distributions are consistent with the MRST parametrization of the ISB of
valance- and sea-quark distributions. Moreover, we find that the
correction to the NuTeV anomaly is in an opposite direction, so the
NuTeV anomaly cannot be removed by isospin symmetry breaking in the
chiral quark model. However, its influence is remarkable and should
be taken into careful consideration. Therefore, it is important to do
more precision experiments and careful theoretical studies on
isospin symmetry breaking.

\section*{Acknowledgement}

This work is supported by the National Natural Science Foundation of
China (grant numbers.~10721063, 10975003, 11035003), and National Fund for
Fostering Talents of Basic Science (grant numbers. J0630311, J0730316). It is also supported by Principal Fund for Undergraduate Research at Peking University.

\end{document}